\documentstyle[aps]{revtex}

\begin{document}

\title{
\rightline{\small accepted in  Phys. Rev. Letters}
    Multiresolution wavelet analysis  of heartbeat 
    intervals discriminates healthy
    patients from those with cardiac pathology
           }

\author{
    Stefan Thurner,$^1$ Markus C. Feurstein,$^1$   
                     and Malvin C.~Teich$^{1,2}$  \\
    $^1$ {\it Department of Electrical and  Computer Engineering,} 
    {\it Boston University, } {\it  Boston, Massachusetts 02215, USA} \\
    $^2$ {\it Departments of Physics, Biomedical Engineering,  and } 
    {\it Cognitive \& Neural Systems, Boston University,}
    {\it   Boston,  Massachusetts 02215, USA} \\
  }

\maketitle

\begin{abstract}
\noindent
We applied multiresolution wavelet analysis to the sequence of times 
between human heartbeats (R-R intervals) and  have found a
scale window, between 16 and 32 heartbeats, over which 
the widths of the R-R wavelet coefficients fall into disjoint 
sets for normal and heart-failure patients. This has enabled us 
to correctly classify every patient in a 
standard data set as either belonging to the heart-failure or normal 
group with 100\% 
accuracy, thereby providing a {\it clinically} significant measure 
of the presence of heart-failure from the R-R intervals alone.
Comparison is made with previous approaches, which have  
provided only  {\it  statistically} significant measures. 
\\ 
PACS number(s): 87.10.+e, 87.80.+s, 87.90.+y 
\\
\end{abstract}

\noindent
Multiresolution wavelet analysis 
\cite{DOBE92,MALL89,MEYE86,ALDR96,METN97} has proved to be a useful technique 
for analyzing  signals at 
multiple scales, even  in the presence of nonstationarities 
which often obscure such signals \cite{ARNE95,TEIC96}. 
The sequence of times between human heartbeats (R-R intervals) 
is a prototype of a nonstationary  
time series that  carries information    
about the state of cardiovascular  health of the patient \cite{KITN82,BASS94}. 

By projecting this sequence into a wavelet space, a new set of 
variables is obtained, whose statistics allow us, for the first time, to 
correctly  classify every patient in a standard data set as  
either heart-failure or normal, with 
100\% accuracy. 
It is clear from our results that the R-R intervals 
alone suffice as a measure for the presence  of heart-failure; the full 
electrocardiogram in not required. 
This remarkable result arises from the ability of multiresolution 
analysis to simultaneously and compactly 
monitor multiple time scales and thereby to expose 
a hitherto unknown scale window (between 16 and 32 heartbeats) 
over which the widths of the R-R 
wavelet coefficients fall into disjoint  sets for normal and 
heart-failure patients. 
The emergence of this particular scale window should help shed 
light on the underlying dynamics of cardiovascular function \cite{BASS94}. 
Previous approaches \cite{PENG93,PENG95,GAND97}, even 
those that  have made use of wavelets \cite{IVAN96}, 
have been successful only in providing a 
{\it statistically} significant measure, rather than the
{\it  clinically} significant one we have developed.
The analysis method we have used is applicable to a 
wide variety of nonstationary physical and biological signals, 
regardless of whether the underlying fluctuations have stochastic origins  
or arise from nonlinear dynamical processes. 

The  series of  intervals between 
adjacent heartbeats $\tau_i$ 
(known as R-R or interbeat intervals in cardiology; see Fig. 1a and 1b) 
is thought to result from a complex    
superposition of  multiple physiological processes at their 
respective characteristic time scales \cite{BASS94}. 
The object of this letter is to demonstrate that is is possible, without 
any {\it a priori} knowledge of the physiological time scales or 
underlying heart dynamics, 
to determine a range of scales over which a statistic of the wavelet 
coefficients permits each heart-failure and normal patient to be correctly 
categorized. 

Scale-dependent 
statistics are constructed by  transforming  
the discrete-time sequence of R-R intervals 
$s=\{\tau_i\} $   \cite{data} 
into a space of wavelet coefficients. 
One can think of the transformed signal  in terms of a landscape over  a 
two-dimensional plane whose axes are  interbeat-interval number  
$i$ and  scale $m$ (see Fig 1c). 
Smaller scales correspond to more rapid variations and therefore to 
higher frequencies. The height is 
the value of the corresponding wavelet coefficient. 
With such a three-dimensional construct  
it is possible to trace the relative importance of 
different scales as the heartbeat sequence  proceeds. 
Technically the coefficients are obtained 
by carrying out a discrete wavelet transform (DWT) \cite{DOBE92,ALDR96}
\begin{equation}
W^{{\rm wav}}_{m,n} (s)= 2^{-m/2}  \sum_{i=1}^{M} \tau_i \psi (2^{-m} i- n)
\, ,
\end{equation}
where the scale variable $m$ and the translation variable  $n$ are
integers, and $M$ represents the total number of R-R intervals
analyzed. The discrete wavelet transform is evaluated at the points $(m,n)$ 
in the scale--interval-number plane.

We have carried out this transformation using  a broad range of  
orthonormal, compactly supported analyzing wavelets. 
We present results for both 
Daubechies 10-tap and Haar wavelets; similar results were obtained  using 
other wavelets.  
Orthogonality in the DWT provides  that the information represented  at a certain
scale $m$ is disjoint from the information at other scales.
Because certain wavelets  $\psi$ 
have vanishing moments, polynomial trends in the signal 
are automatically eliminated in the process of wavelet transformation 
\cite{ARNE95,TEIC96,ABRY96}. 
This is salutatory in  the case of the heartbeat time series, 
as is evident from the trends apparent in Fig. 1b, 
which are eliminated by the wavelet transformation   
shown in Fig. 1d.

Since the signal $s$ fluctuates in time, so too does the 
sequence  of wavelet coefficients at any given 
scale, though its mean is zero \cite{DOBE92}.
In Fig. 2 we present   
histograms  of the wavelet coefficients at six different scales 
($m=$ 2 - 7) for a  
normal (left) and a  heart-failure patient (right). 
The wavelet coefficients for the heart-failure patient evidently exhibit 
substantially reduced variability, particularly at intermediate scales. A 
natural measure for this variability is the wavelet-coefficient standard 
deviation, 
as a function of scale:    
\begin{equation}
\sigma_{{\rm wav}}(m) = \left[  \frac{1}{N-1} 
                \sum_{n=1}^N (W^{{\rm wav}}_{m,n}(s) - 
                \langle W^{{\rm wav}}_{m,n}(s) \rangle )^2 
                \right] ^{\frac{1}{2}} ,
\end{equation}
where  $N$ is the number of wavelet coefficients at a given scale $m$ 
($N=M/2^m$).

The principal results of this paper are displayed in Fig. 3, where 
$\sigma_{{\rm wav}}$ is plotted vs. scale 
($1 \leq m \leq 10$) for all of 
the 27 patients (open and solid circles) \cite{data}, using the 
Haar wavelet (left) and the Daubechies 10-tap
wavelet (right). 
At scales 4 and 5, corresponding to $2^4$ - $2^5$ = $16$ - $32$ heartbeats,  
$\sigma_{{\rm wav}}$ serves to completely separate the two classes of 
patients for both types of wavelets (white regions), 
thereby providing a clinically significant 
measure \cite{SWET88,TURC96} of the presence of heart failure 
with 100\% sensitivity at 100\% specificity 
(such that all normals are so identified). One can do no better. 
Though it has been previously shown that complete separation 
can be achieved using 
heart{\it rate} analysis \cite{TURC96,TURC93}, 
this is the first instance that we know of, in which the 
R-R intervals can be used as a definitive determinant 
of the presence of a heart disorder in an individual patient. 
Both at smaller, and at larger scales, there are multiple overlaps 
of the heart-failures and the normals,
though the measure certainly remains statistically 
significant over all scales shown. Similar results are obtained for other 
analyzing wavelets.

The results indicate that healthy patients exhibit greater fluctuations than
those afflicted with heart failure over a time scale of 16-32 heartbeats
(roughly 0.2 to 0.5 minutes).  This also appears to apply
for sudden cardiac death (SCD) as illustrated by the
open squares in Fig. 3. It is tempting to ascribe the
physiological origin of this window to
baroreflex modulations of the sympathetic or
parasympathetic tone, which lie in the range 0.04 to 0.09 Hz (0.2 to 0.5
minutes), but we do not believe that this is correct. Rather, we
expect that this window
likely has its origin in the intrinsic behavior of the heart itself.
It will be important to carry out a thorough study in which
our multiresolution wavelet-analysis technique is applied
to the R-R intervals from transplanted hearts to assess the role that
the autonomic system might play in heartrate variability.

It is useful to tease apart the roles played by
the {\it magnitudes} $\tau_i$ of the interbeat intervals
and their {\it ordering} in 
achieving this complete separation. The effects of the former continue to reside
in the randomly reordered (shuffled) sequence of R-R intervals; however
information about the ordering is removed in this surrogate
data set \cite{TURC96}. We therefore calculate the 
standard deviation $\sigma^{{\rm shuf}}_{{\rm wav}}$ 
for all 27 heartbeat time series after shuffling the R-R intervals. 
[The first $70\,000$ interbeat intervals were selected for analysis
after the entire data sets (see Table 1 in Ref. \cite{TURC96})
were shuffled.] The results are shown in Fig. 4a. 
It is clear that the two classes of patients are 
no longer  completely separated; 
three heart-failure patients fall among the normals at all scales, 
yielding a sensitivity of 80\% at a specificity of 100\%. 
Comparison with Fig. 4b shows that the shuffled-wavelet  result is essentially 
identical to that obtained by using the standard deviation 
$\sigma_{{\rm int}}$ of the interbeat-interval 
histogram (IIH) obtained from these data sets, 
a measure that has long been used in cardiology \cite{WOLF78,BIGG87}. 
The concurrence in sensitivity displayed in Figs. 4a and 4b is not 
accidental; the shuffled interbeat intervals essentially comprise 
a renewal process so that the 
IIH contains all of the  available information.  
All dependencies among intervals, and  
therefore  long-term correlations,
are removed from the shuffled surrogate data  
\cite{TURC96}, leaving behind only short-term information.
Indeed, for the Haar analyzing wavelet, 
$\sigma_{{\rm wav}}(m=0)$ (in the absence or in the presence of 
shuffling) is analytically identical to 
$\sigma_{{\rm int}}$. 


The ordering of the interbeat intervals gives rise to scaling 
behavior, as is evident from a comparison of Figs. 3 and 4a.
For normal patients (open circles, solid lines) 
the straight-line behavior in Fig. 3 indicates that
approximate scaling is maintained across all scales whereas
for heart-failure patients
(filled circles, dashed lines) the relatively flat nature of the curves
in the region $m \leq 3$ indicates that $\sigma_{{\rm wav}}$ 
is essentially scale independent in this region. 

The distinction can be examined quantitatively by calculating the
average scaling exponents $\alpha$ \cite{slope} 
in the two ranges ($1 \leq m \leq 3$ and
$3 \leq m \leq 10$), for both classes of data.
The results are reported in Table 1, 
along with the average scaling exponents determined across
the entire scaling region examined ($1 \leq m \leq 10$). 
It is clear that for the
12 normal patients, the value of $\alpha$ is quite insensitive 
to the range over which
it is estimated; it is nearly the same in the two subregions 
as it is in the entire range. For the 15 heart-failure patients, on
the other hand, the scaling exponent estimated in the region 
$1 \leq m \leq 3$ is dramatically lower than that estimated in the
region $3 \leq m \leq 10$. Over the range of larger scales, the values of the
scaling exponents $\alpha$ provided by our wavelet measure (see Table 1),
are in good accord with typical values $\delta$ provided by
the interval-based periodogram at low frequencies
(see Table 1 in Ref. \cite{TURC96}), as expected.
We conclude that scaling persists 
across a broader range for normal patients than it does for heart-failures,
as is visually evident in Fig. 3.

These observations lead us to consider a heart-failure index determined by
the difference of these scaling exponents:
$\Delta=\alpha (3 \leq m \leq 10)-\alpha (1 \leq m \leq 3)$. 
Evaluating $\Delta$ we find 
that two heart-failures fall among the normals, corresponding to a 
sensitivity of 87\% at a specificity of 100\%. Thus considering only the
scaling information in $\sigma_{{\rm wav}}$, while ignoring the 
magnitude differences for normals and heart-failures 
associated with short-term information 
(as illustrated in Fig. 4a), fails to give rise to 
complete separation.  

Over the years, using this same collection of data, a number of 
measures based on scaling have been evaluated for their accuracy 
in discriminating between normal and heart-failure patients. 
Peng {\it et al.} \cite{PENG93} examined the correlation properties 
of the  heartbeat-interval increments  $I_i=\{\tau_{i+1}-\tau_i \}$, 
and obtained the exponent of the associated power-law spectrum,    
which they denoted as $\beta$.  
It was shown subsequently \cite{TURC96} that this measure 
is isomorphic to the exponent $\delta$ of the interval-based spectrum
\cite{GORD81}   
at low frequencies \cite{KOBA82} and therefore reveals
only long-term correlations. 
We have calculated $\beta$ for all 27 data sets and present the
results in Fig. 4c. This
measure results in 7 heart-failures among the normals, 
yielding a sensitivity of 53\% at 100\% specificity, so that it is not 
particularly successful in
discriminating the two classes of patients.
The interbeat-interval standard deviation $\sigma_{{\rm int}}$ shown
in Fig. 4b (sensitivity of 80\% at 100\% specificity) performs 
significantly better. 
In a subsequent paper, Peng and collaborators 
\cite{PENG95} constructed a "detrended fluctuation measure" 
which turns out to yield approximately 80\% 
sensitivity at 100\% specificity for the 27 data sets. Most recently, 
this same group introduced a so-called "scaling instability index" 
\cite{GAND97}, which achieved a sensitivity of 
71\% at 100\% specificity for the 25 (of the 
same 27) data sets that they reported, as shown in Fig. 4d. 
The performance of both of these latter measures is
therefore comparable to that achievable with the interbeat-interval standard 
deviation measure $\sigma_{{\rm int}}$ introduced by Wolf et al. long ago
\cite{WOLF78}.
As far as we are aware, no group has been able to 
successfully separate this standard collection of data with 100\% 
sensitivity at 100\% specificity using scaling measures alone.
 

We conclude that our multiresolution approach succeeds not only because 
it eliminates trends in a mathematically acceptable way, 
but also because it crisply reveals a range of 
scales over which heart-failure patients differ from normals, both in 
short- and long-term heartbeat behavior. 
In contrast, interbeat-interval measures 
reflect only short-term behavior, whereas scaling measures 
reflect only long-term behavior. 
Our approach should have  wide applicability in the analysis of 
nonstationary processes in the  physical and biological sciences.

\vspace{0.5cm}

\noindent 
We are grateful to S.B. Lowen for valuable comments.



\noindent 

\vspace{2cm}

\begin{figure}
\caption{
      (a) Schematic diagram  of  an electrocardiogram  segment, 
          showing the beat occurrence times $t_i$ and the interbeat (R-R) intervals $\tau_i$.
      (b) Series of interbeat intervals $\tau_i$ versus  interval number $i$ for a typical 
          normal patient (data set 16265). (Adjacent values of the interbeat interval
          are connected by straight lines to facilitate viewing.) 
          Substantial trends are evident.  
      (c) 3-D  representation of the wavelet coefficient $W$ as
          a function of scale ($1 \le m \le 10$) and interval number $i$, 
          over a portion of the data set, using 
          a Daubechies 10-tap analyzing wavelet.
      (d) Wavelet coefficient at three scales ($m$ = 2, 4, 8) for the data set 
          illustrated in Fig 1b. The trends in the original interbeat-interval 
          time series are removed by the wavelet transformation.
    }
\end{figure}
\begin{figure}
\caption{  Typical histograms of the interbeat-interval wavelet coefficients  
           for a  normal patient (data set 16265, left panels) and for a  
           heart-failure patient  (data set 6796, right panels), for different 
           values of the scale $m$, using a  Daubechies 
           10-tap wavelet.  The ordinate is the number of wavelet coefficients 
           $N$. Heart-failure patients tend 
           to have narrower distributions, particularly at intermediate scales. 
         }
\end{figure}
\begin{figure}
\caption{Wavelet-coefficient  standard deviation $\sigma_{{\rm wav}}$ versus scale $m$ 
         for the standard  27 data-set collection 
         (12 normals and 15 heart-failures indicated by circles) [14], 
         using the Haar 
         wavelet (left) and the  Daubechies 10-tap wavelet (right). 
         Complete separation of the two groups  is 
         achieved at scales 4 and 5, corresponding to $2^4$ - $2^5$  
         heartbeats. 
         Results for an SCD patient (white squares), using the
         same number of interbeat intervals, also exhibit low variability.
         The outcome is similar for both analyzing wavelets, though 
         the separation is slightly more pronounced for the Daubechies wavelet, 
         most likely because of its greater number of 
         vanishing moments.
}
\end{figure}

\begin{figure}
\caption{ (a) Wavelet-coefficient standard deviation 
              $\sigma^{{\rm shuf}}_{{\rm wav}}$ versus scale $m$ 
              for the standard  27 data-set collection [14]
              after random reordering of the 
              R-R intervals (shuffling), using  
              a  Daubechies 10-tap wavelet. 
              Partial separation of the two groups is achieved; 
              three heart-failure patients fall among the normals at all scales, 
              yielding a sensitivity of 80\% at a
              specificity of 100\%. 
          (b) Standard deviation of the interbeat intervals $\sigma_{{\rm int}}$ 
              for all 27 data sets. 
              This measure was first used by Wolf {\it et al.} [19]. 
              Three heart-failure patients fall 
              among the normals, corresponding to 
              a sensitivity of 80\% at 100\% specificity. 
          (c) Exponent $\beta$ of  the heartbeat-interval 
              increment-process spectrum for all 27 data sets; this measure 
              was used by Peng {\it et al.} [10].
              Seven heart-failures fall among the normals, 
              yielding 53\%  sensitivity at 100\% specificity. 
          (d) The ``scaling instability index,'' recently introduced 
              by Peng and collaborators  
              [12], and reported by them for 25 of the same 27 data sets. 
              Four heart-failures fall among the 
              normals, yielding 71\% sensitivity at 100\% specificity, 
              still falling  
              short of our multiresolution  results shown in Fig. 3 
              which yield  100\% sensitivity 
              at 100\% specificity. 
}
\end{figure}

\begin{table*}
\begin{tabular}{ l c c c c}
\multicolumn{5}{c} {\bf Interbeat Intervals  (27 Patients) } \\
\hline
\multicolumn{1}{l}{Class of Patients} & 
\multicolumn{4}{c}{Results} \\
\hline
\multicolumn{1}{c}{} & 
\multicolumn{1}{c}{$\langle \tau_i \rangle$} & 
\multicolumn{1}{c}{$\alpha  (1 \leq m \leq 10)$} & 
\multicolumn{1}{c}{$\alpha  (1 \leq m \leq 3)$ } & 
\multicolumn{1}{c}{$\alpha  (3 \leq m \leq 10)$ }\\
\hline
\multicolumn{1}{l}{Normal (12)} &
\multicolumn{1}{ c } {   $0.79 \pm 0.08$} & 
\multicolumn{1}{ c } {   $1.23 \pm 0.13$} & 
\multicolumn{1}{ c } {   $1.40 \pm 0.37$} &
\multicolumn{1}{ c } {   $1.22 \pm 0.11$} \\
\multicolumn{1}{l}{HF (15)} &
\multicolumn{1}{ c } {   $0.67 \pm 0.13$} & 
\multicolumn{1}{ c } {   $1.35 \pm 0.22$} & 
\multicolumn{1}{ c } {   $0.26 \pm 0.60$} &
\multicolumn{1}{ c } {   $1.57 \pm 0.17$}  \\
\end{tabular}
\caption{ Mean interbeat intervals 
          and estimated scaling exponents $\alpha$ 
          (with their standard deviations) for the 12 normal and 
          15 heart-failure patients, collected in two separate groups. 
          Scaling-exponent estimates 
          are evaluated over the whole range of $m$, as well as over two 
          smaller scaling subregions: $1\leq m\leq 3$ and $3\leq m\leq 10$. 
          The scaling behavior is significantly different for 
          the two groups of patients. The difference of the scaling 
          exponents in the two subregions,  
          $\Delta=\alpha (3 \leq m \leq 10)-\alpha (1 \leq m \leq 3)$, 
          leads to a sensitivity of 87\% at 100\% specificity. 
         }
\label{tab1}
\end{table*}


\begin{thebibliography}{99}
\bibitem{data}   
The R-R recordings were drawn from 
the Beth-Israel Hospital (Boston, MA) Congestive Heart-Failure
Database comprising 12  records
from normal patients  (age: 29-64 years, mean 44 years) 
and 15 records from severe congestive heart-failure patients (age 22-71 years, mean 56 years). 
The recordings, which form a standard database for evaluating the merits of 
various measures for identifying heart-failure, 
were made with a Holter monitor,  digitized at a fixed value of 250 Hz.
Three of the 15 heart-failure patients also suffered  from atrial fibrillation. 
Detailed characterization of the data sets is presented in Table 1 of Ref. \cite{TURC96}.
We used the first $M =  70\,000$ interbeat intervals (total time 
duration $T \approx $ 20 h) of each of these records.
\bibitem{slope}
The scaling exponent $\alpha$ is computed from an individual
wavelet-coefficient standard-deviation curve (as
shown in Fig. 3) by calculating the slope of its square 
(rendering it a variance so that it corresponds to other standard
scaling-exponent measures) on a base-10 log-log plot. Thus
$\alpha = [d({\rm log_{10}}\sigma_{{\rm wav}}^{2}(m))]/[d({\rm log_{10}}2^m)] =
[2/{\rm log_{10}}2][d({\rm log_{10}}\sigma_{{\rm wav}}(m))/dm]$.
\end{thebibliography}
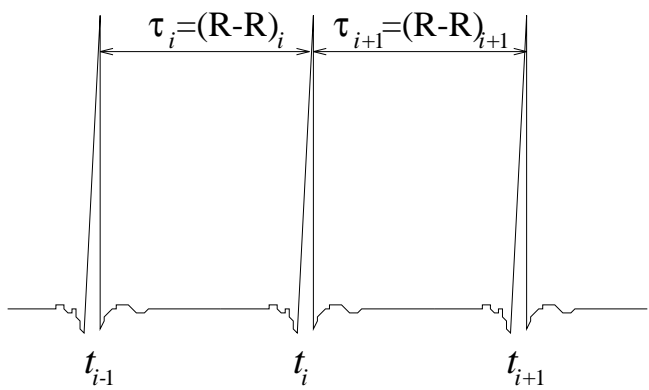
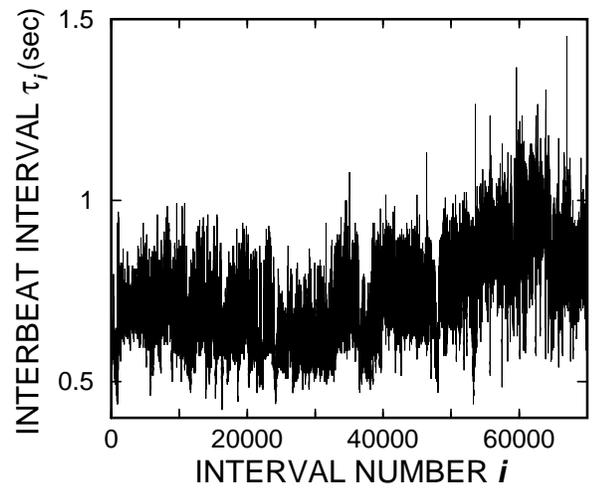
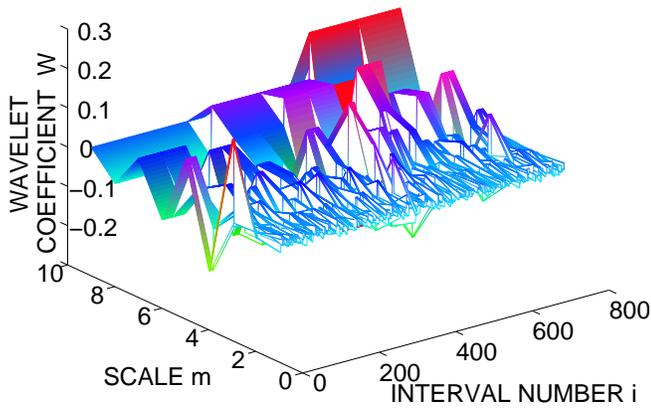
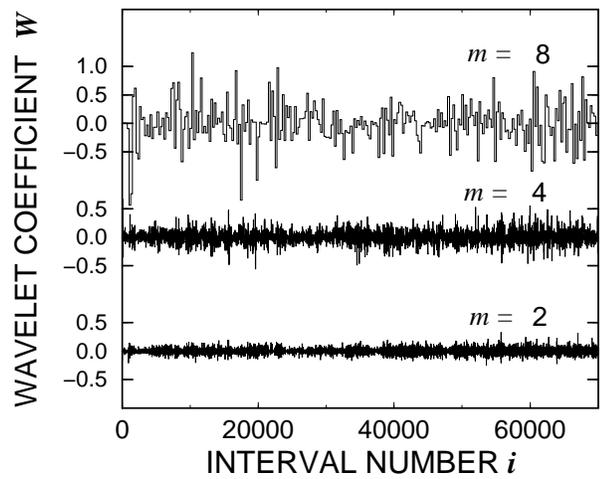
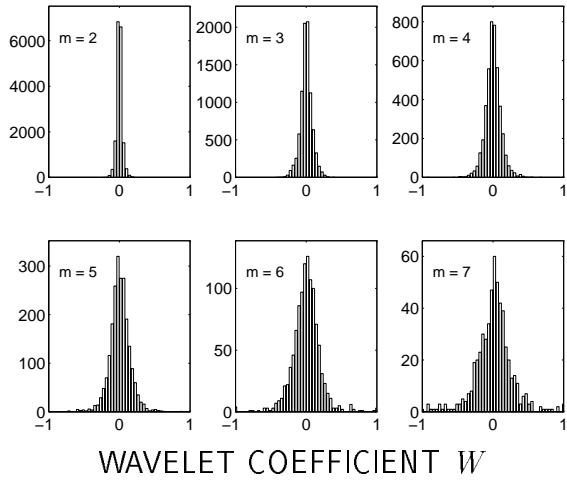
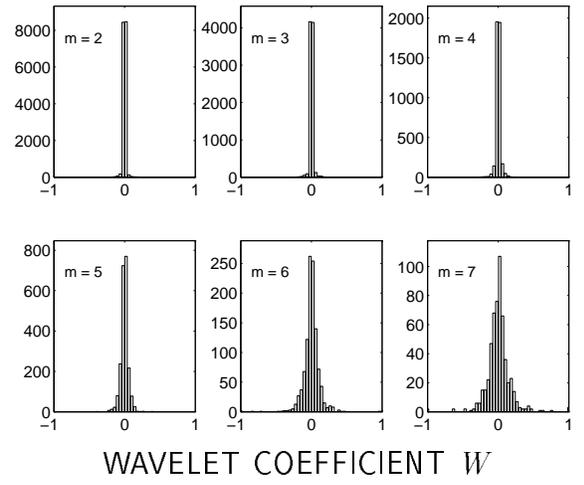
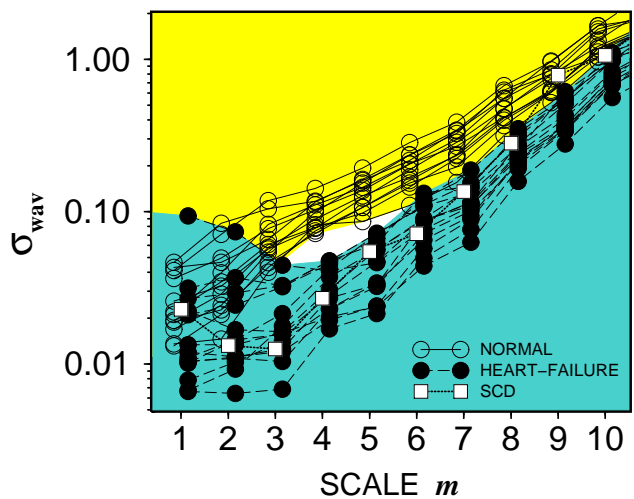
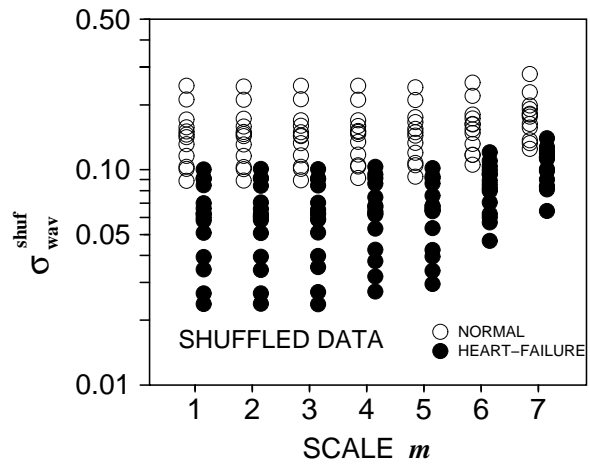
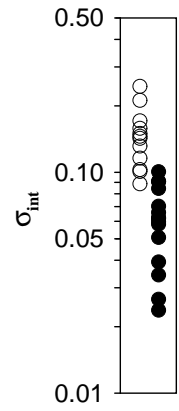
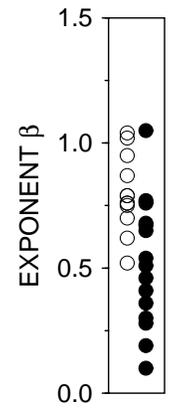
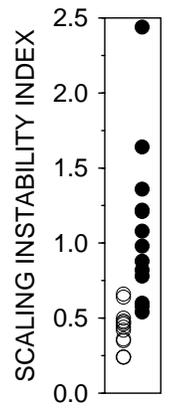
\end{document}